\documentclass[twocolumn,aip,rsi,reprint,numerical,superscriptaddress]{revtex4-1}
\usepackage{times}
\usepackage{mathptm}
\usepackage{amsmath}
\usepackage{graphicx}
\usepackage{color}

\begin{document}
\title{
The molecular yo-yo method: Live jump detection improves throughput of single-molecule force spectroscopy for out-of-equilibrium transitions}
\author {A. H. Mack}
\thanks{These authors contributed equally to this work.}
\affiliation{Integrated Graduate Program in Physical and Engineering Biology, Yale University, New Haven CT 06511}
\affiliation{Department of Applied Physics, Yale University, New Haven CT 06511}
\author{D. J. Schlingman}
\thanks{These authors contributed equally to this work.}
\affiliation{Integrated Graduate Program in Physical and Engineering Biology, Yale University, New Haven CT 06511}
\affiliation{Department of Molecular Biophysics and Biochemistry, Yale University, New Haven CT 06511}
\author{M.  Kamenetska}
\affiliation{Department of Molecular Biophysics and Biochemistry, Yale University, New Haven CT 06511}
\affiliation{Department of Physics, Yale University, New Haven CT 06511}
\author{Robert Collins}
\affiliation{Department of Molecular Biophysics and Biochemistry, Yale University, New Haven CT 06511}
\author{L.  Regan}
\affiliation{Integrated Graduate Program in Physical and Engineering Biology, Yale University, New Haven CT 06511}
\affiliation{Department of Molecular Biophysics and Biochemistry, Yale University, New Haven CT 06511}
\affiliation{Department of Chemistry, Yale University, New Haven CT 06511}
\author{S. G. J. Mochrie}
\affiliation{Integrated Graduate Program in Physical and Engineering Biology, Yale University, New Haven CT 06511}
\affiliation{Department of Applied Physics, Yale University, New Haven CT 06511}
\affiliation{Department of Physics, Yale University, New Haven CT 06511}

\date{\today}


\begin{abstract}
By monitoring multiple molecular transitions,
force-clamp and trap-position-clamp methods have led to precise determinations of the free energies
and free energy landscapes for molecular states populated in equilibrium at the same or similar forces.
Here, we present a powerful new elaboration of the force-clamp
and force-jump methods, applicable to transitions far from equilibrium.
Specifically, we have implemented a live jump detection and force-clamp
algorithm that intelligently adjusts and maintains the force on a single molecule
in response to the measured state of that molecule.
We are able to collect hundreds of individual molecular transitions at different forces,
many times faster than previously, permitting us to accurately determine
force-dependent lifetime distributions and reaction rates.
Application of our method to unwinding and rewinding the nucleosome inner turn,
using optical tweezers
reveals experimental lifetime distributions that comprise a statistically-meaningful number of transitions,
and that are accurately single exponential.
These measurements significantly reduce the error in the previously measured
rates, and demonstrate the existence of a single, dominant free energy barrier at each force studied.
A key benefit of the molecular yoyo method for nucleosomes is that it reduces as
far as possible the time spent in the tangentially-bound state, which minimizes the loss of
nucleosomes by dissociation.
\end{abstract}

\maketitle
\setcitestyle{round}
\section{INTRODUCTION}
Single-molecule studies of biomolecules and their complexes have 
led to important new insights into the molecular mechanisms underlying many
biological processes. Single-molecule force spectroscopy (SMFS), in particular,
has proven a powerful method by which to study the mechanochemistry of molecular motors,
including kinesins \cite{Block1990,svoboda1993direct,Kuo1993},
myosins \cite{Molloy1995,Wolenski1995,Veigel1999},
polymerases \cite{Wang1998,Forde2002,Abbo2005,Herbert2006,Bintu2011},
helicases
\cite{Perkins2004,Dumont2006,Johnson2007},
chromatin remodellers \cite{Lia2006, Zhang2006,Sirinakis2011},
and the ribosome
\cite{Wen2008,Qu2011}.
SMFS has also transformed our understanding of nucleic acid \cite{Liphardt2001,Woodside2006,Woodside2008,LaPorta2011}
and protein folding
\cite{Rief:1997,Marszalek:1999,Li:2001,Best:2002,Fernandez:2004,Cecconi2005,Brujic2006,Gebhardt2010,Gao2011,Yu2012},
and biological assembly processes, such as chromatin compaction
\cite{Cui2000,Bennink2001,Brower2002,Brower2005,Gemmen2005,Pope2005,Mihardja2006,Kruithof2009b,Kruithof2009,Hall2009,Simon2011,Mack2012a,Molla2012}.
%
In all of these cases, the key questions are: What microscopic states are populated at a given force? What are the characteristics of the force-dependent transitions among these states, including what is the distribution of lifetimes for each transition and what is the mean rate of each transition? These quantities determine the relevant free energy
landscape of the biomolecules under study, and their diffusion constants for motion within this landscape. In practice, however, the level of detail resolvable in the free energy landscape may be limited by the number of molecular events that can feasibly be observed.

 To-date, the most detailed characterizations of biomolecular free energies and free energy landscapes have been obtained in optical-tweezers-based force-clamp
 \cite{Simmons1996,Visscher1998,Oberhauser:2001,Fernandez:2004,Brujic2006}
 and trap-position-clamp
 \cite{Gebhardt2010,LaPorta2011}
 experiments in which multiple molecular states are
 in equilibrium with each other at the same or nearly the same force.
 In such situations, 
 it is often possible  to observe
 hundreds of transitions back and forth among the different molecular states involved.
 Because of the large number of transitions,  the transition rates, and hence the free energies of the states,
 may be accurately determined
  \cite{Liphardt2001, Cecconi2005, Woodside2006, Mihardja2006, Woodside2008,Gebhardt2010,LaPorta2011, Gao2011,Yu2012}.
 By contrast,
 for molecular states that are not in equilibrium -- 
 %
 that is, for molecular states for which spontaneous transitions back and forth are not observable on experimental timescales at a single force
 -- it is more challenging to collect a statistically large-enough data set to be able to
 determine the transition rates accurately.
 Most commonly under these circumstances, researchers carry out repeated measurements of the molecular
 force-versus-extension curve \cite{Rief:1997,Mihardja2006,gebhardt2010full,strunz1999dynamic} in order to determine the distribution of transition forces. To increase the rate of data acquisition in force-versus-extension measurements, tandem arrays of independent, identical molecules are often used
 \cite{Rief:1997,Marszalek:1999,Li:2001,Best:2002}.
 Nevertheless, each force-versus-extension curve contains a limited number of transitions, and several practical limitations, such as surface sticking, tether rupture, protein dissociation, {\em etc.} mean that each molecule is only measurable for a finite period of time. Moreover, a fundamental limitation of the force-versus-extension approach, compared to force-clamp and trap-position-clamp methods, is that the force-dependent lifetime distributions and transition rates are not measured directly. Instead, model-based approaches are required to determine the transition rates from the measured distribution of transition forces,
 in which it is usually assumed that the transition lifetimes are exponentially distributed at all forces
\cite{dudko2007extracting,Dudko2006}.
 The force-jump method,
 in which a force-clamp alternates periodically
 between a high-force and a low-force  \cite{Li2006a,Kaiser2011,Elms2012},
overcomes a number of these difficulties.
 However,  because the force-clamps must be maintained for
 long enough to ensure that the transitions have taken place, this method is
 not efficient. In addition, because the  force must be maintained at a high value for an extended period of time,
the force-jump method is especially vulnerable to tether rupture and other irreversible damage.
 
The purpose of this paper is to present a powerful new elaboration of the force-clamp
and force-jump methods,
which we call ``the molecular yo-yo method'', that permits the efficient determination
of lifetime distributions and transition rates for single molecules, even for molecular states that are not in equilibrium.
The key innovation of
the molecular yo-yo method to implement a live state-detection and force-clamp  algorithm,
that intelligently adjusts and maintains the force on a single molecule in real-time,
in response to the measured state of the molecule. We describe two simple to implement live jump detection algorithms.
The molecular yo-yo method is broadly applicable to out-of-equilibrium molecular transitions of all sorts.
Here, to showcase the usefulness of this method, we report 
its
implementation
in experiments that seek an improved characterization of the unwinding and
rewinding transitions of the nucleosome inner turn,
which are out-of-equilibrium at near-physiological salt concentrations
\cite{Cui2000,Bennink2001,Brower2002,Brower2005,Gemmen2005,Pope2005,Mihardja2006,Kruithof2009b,Kruithof2009,Hall2009,Simon2011,Mack2012a,Molla2012}.

\begin{figure}[t!]
	\begin{center}
		\includegraphics[trim= 0.0cm 1.5cm 0cm 1.9cm, clip=true,width=3.2in]{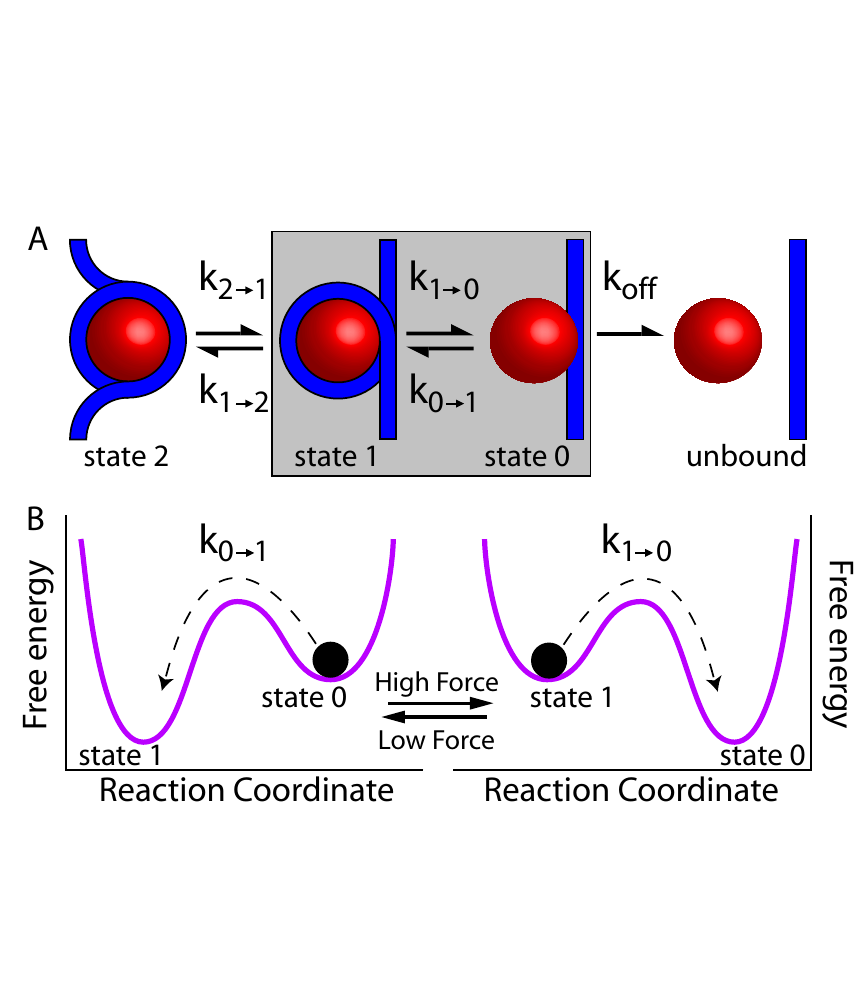}
	\caption{	(A) Schematic of the four microstates for 
nucleosome unwinding/rewinding
\cite{Brower2002}.
Histones are red.
DNA is blue.
For  state 2, the nucleosome is fully wrapped by nearly two turns of DNA.
For state 1, the outer turn is unwrapped, but the inner turn is wrapped.
For state 0, both the outer and inner turns are unwrapped,
but the histone octamer remains bound to the DNA.
Finally, the histones may be unbound from the DNA. The shaded box encompasses the states and
transitions probed in this paper.
(B) Schematic illustration of the application of our molecular yo-yo method to the free energy landscape for states not in equilibrium. 
The rate of transitions from state 1 to state 0 is non-negligible only at high force;
the rate of transitions from state 0 to state 1 is non-negligible only at low force.
 Following every transition, our algorithm detects the molecular state and adjusts the force so that the reverse transition can take place.
 By toggling between the high and low force configurations we are able to rapidly measure many transitions from a single molecule.
}
\label{Fig0}
	\end{center}
\end{figure}

The  relevant states for  nucleosome unwinding and rewinding
are illustrated in Fig.~\ref{Fig0}A  \cite{Brower2002}.
In this paper, we probe the transition from state 1 to state 0 as shown in Fig.~\ref{Fig0}B.
Although states 1 and 0 are out of equilibrium at any force, nevertheless, using the molecular yo-yo method,
we have been able to collect hundreds of molecular transitions,
permitting us to accurately determine force-dependent lifetime distributions
and force-dependent reaction rates for unwinding and rewinding the nucleosome inner turn.
The molecular yoyo method is especially valuable for studies of nucleosomes
because it reduces as far as possible the time spent in the unwound state (state 0),
correspondingly reducing the loss of nucleosomes by dissociation in a given period of time.
Using this new method, we found that at each force studied, the lifetime distributions are  well-described as a single exponential,
indicating that the free energy landscapes relevant to winding and unwinding are each dominated by a single free energy barrier. 
We also demonstrate that the unwinding rates for tethers containing 2, 4 and 8 nucleosomes are accurately 2-, 4-, and 8-fold faster, respectively, than for tethers containing a single nucleosome.
This observation implies that nucleosomes on the same tether unwind independently,
as has been previously assumed but not proven.
Finally, we show that the rates measured using the optical yo-yo method match those determined previously by the more laborious method of making a series of measurements at different forces, using a simple force clamp and force jumping.

{

\section{MATERIALS AND METHODS}

\subsection{DNA preparation and surface attachment}

Segments of DNA containing 4200 base-pairs,
corresponding to a contour length of 1430~nm,
were created from linearized pUC18 plasmids harboring an array of twelve 601 nucleosome positioning sequences (NPSs), a kind gift from Dr. Daniela Rhoades \cite{Lowary1998,Huynh2005}.
To facilitate robust attachments between the DNA and a microscope coverslip and
between the DNA and the optically-trapped bead,
the DNA is labeled with biotin on one end and an amine on the other end. 
To create 
DNA-tethered beads, the amine-labeled DNA end is covalently attached to the
 glass coverslip via a silane PEG N-hydroxysuccinamide linker while
 the biotin-labelled DNA end binds to a streptavidin coated polystyrene bead
 \cite{Schlingman}.
 
\subsection{Histone expression and purification}
pET vectors containing untagged {\em Xenopus} H2A, H2B, H3 or H4 were a kind gift from Dr. Karolin Luger.
Histones are expressed in {\em E. coli} BL21-Gold(DE3), extracted from isolated inclusion bodies in buffer containing 7~M GuHCl, and dialyzed into 8~M urea buffer. Histones are purified first by passage through a Q-sepharose column, and then bound to a Hi-Trap SP column (GE Healthcare Life Sciences), washed with 300 mM~NaCl, and eluted with a step gradient to 600~mM NaCl. Finally, the histones are dialyzed into deionized distilled H$_2$O, lyophilized, and stored at $-80^\circ$C until needed. 

\subsection{Nucleosome reconstitution}
Equimolar ratios of the four core histones are combined in buffer containing 7~M GuHCl and dialyzed into buffer containing 2~M NaCl, resulting in octamer formation  \cite{Luger1999}.
After isolation by gel filtration, octamer is mixed with carrier DNA 
-- ultra pure salmon sperm DNA, sheared to $1000$~bp (Invitrogen) --
 and continuously dialyzed into buffer without salt to form nucleosomes. 
We then assemble nucleosomes {\em in situ} by flowing
a solution of nucleosomes, bound to carrier DNA,
 at 680~mM NaCl,  into our optical tweezers flow cell,
 which
consists of a flow channel cut out of double-sided sticky tape between a microscope coverslip 
and a microscope slide into which are drilled two holes for fluid inlet and outlet, respectively. 
{\em In-situ} nucleosomes exchange between the carrier DNA and the immobilized 601 DNA tether 
ensures occupancy of the 601 sites.
We then flow in a 100~mM NaCl buffer,
in preparation
for optical tweezers measurements.
For experiments discussed here, we prepared arrays of varying numbers of nucleosomes per tether in order to be able to compare rates of unwinding and rewinding with different number of nucleosomes. 

\subsection{Optical trapping instrumentation}
In the optical trapping setup used for these experiments,
the 
 beam from a
1064~nm laser (Ventus IR, Laser Quantum, Stockport, UK) is incident on an
 acousto-optic deflector (AOD) (IntraAction DTD-274HA6), which serves to
optically isolate the laser from the downstream optics,
 and to control the trapping laser power.
Located between the AOD and the microscope objective
(Nikon CFI  $\times 100$, oil immersion, NA~1.25)
is a telescope 
that expands the beam by a factor of three to
ensure that the back pupil of the objective is overfilled, as required for strong trapping.
Beyond the objective, the transmitted laser light is incident upon a quadrant photo-diode (QPD)
(Phresh Photonics SiQu50-M),
located in a plane conjugate to the back focal plane of the microscope condenser lens,
where  variations in the summed intensity of  all four QPD quadrants
are linearly proportional to the displacements of a trapped bead from the center of the trap along the beam direction.
This method of determining bead position -- ``back-focal-plane interferometry'' (BFPI) \cite{GittesOL1998}
-- provides a sensitive measure of the force on the bead,
which is proportional to the bead displacement from the trap center. 

The measurements described in this paper were carried out
using an axial pulling
geometry in which a piezo-electric translation stage moves
the microscope coverslip along the direction of the laser beam,
thus applying tension to a surface-tethered molecule in
the axial direction and maintaining a simple geometry at all extensions
\cite{clapp1999three,neuman2005measurement,Deufel2006,Chen2009,Forns20111765,LaPorta2011}.
To make use of the axial pulling geometry, we have implemented a
new calibration method that allows conversion
from the experimental signals -- stage displacement and scattering intensity -- to 
calibrated values of the molecular extension and applied tension.
Our calibration procedures are fully described in Ref.~\onlinecite{Mack2012rsi}.

Our axial pulling geometry enables straightforward implementation of a
reliable feedback loop that maintains a constant force on the
tether \cite{Mack2012rsi}. 
Specifically, the force measured by the QPD is held constant by adjusting the position of a piezo-electric
stage (NanoMAX311, Thorlabs).
To achieve this force-clamp, we implemented a proportional-integral-derivative (PID) feedback controller
using LabView and Labview-MathScript, which
carries out the conversion from QPD intensity and piezo-stage position to axial force and tether extension, followed by
actuation of the piezo-stage PID control,  at a cycle rate greater than 1000~Hz.
Data acquisition card PCIe-6343 (National Instruments) is used for both acquisition and output.

When a nucleosome unwinds,  the tether length increases by about 25~nm.
Concomitantly, 
the force transiently decreases.
However, within 10~ms, our force clamp has adjusted the trap position to increase the tension to the force-clamp value.
This response time is limited by the
mechanical response
of the microscope stage, not by the computation time of our conversion algorithm.
In fact, we can flexibly program an arbitrary sequence of forces versus time,
jumping out, if necessary, at programmable break points.
This flexibility enables the molecular yo-yo method.


In order to sustain meaningful measurements on the same molecular construct
for extended periods of time, it is necessary to
dynamically correct for any drift in the position of the piezo-stage  and in the laser intensity.
Since such drift is slow, it is satisfactory to apply  a drift correction procedure every $\sim$200~s,
which we do automatically.
Specifically, stage drift is corrected for by measuring the position at which the bead contacts the coverslip, 
 permitting us to update the piezo-stage calibration correspondingly.
By holding the bead against the stage and measuring for 100~ms,
we are able to establish the bead-coverslip separation to within about 2~nm.
To correct for possible laser intensity drift, the force zero is established by placing the bead close to,
but not in contact with, the coverslip, so that the tether has a very low extension, and the corresponding force is negligible.
This procedure establishes the force to within 0.1~pN.

\begin{figure*}[t!]
\begin{center}
\includegraphics[trim=0cm 1.4cm 0cm 1.75cm, clip=true, width=4.4in]{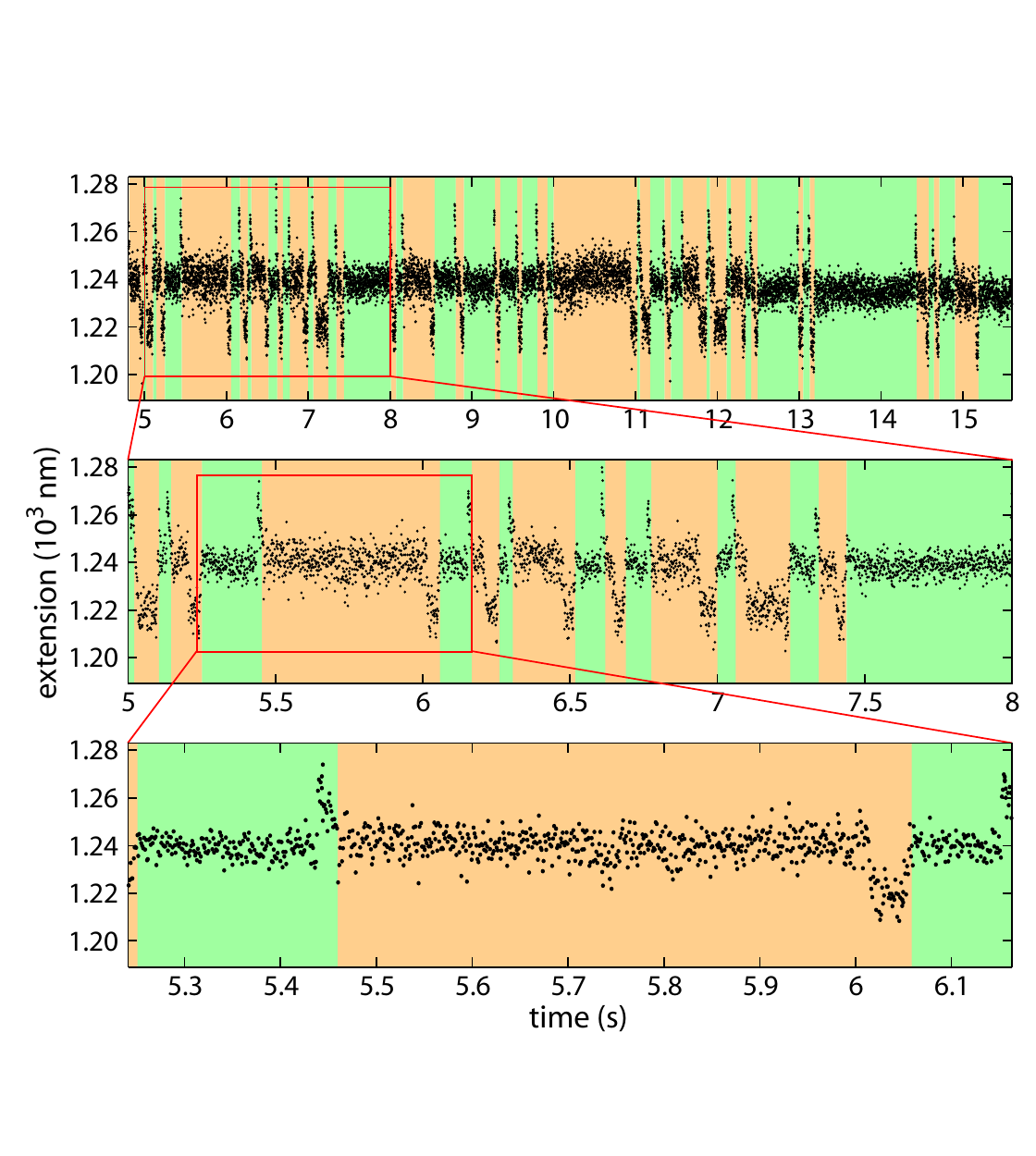}
\end{center}
\caption{Unwinding and rewinding the nucleosome inner turn using the
optical yo-yo method.
(A) Extension versus time while a single nucleosome is repeatedly unwound at an unwinding force of
10.3~pN and rewound at a rewinding force of 3.8~pN over a period of about 11~s.
Extension data obtained at the unwinding force  is shown with a green
background while the extension data obtained at the rewinding
force is shown with an orange background.
(B) Extension versus time, plotted over a restricted time range,
showing four unwinding events from (A).
(C) Extension versus time plotted for a further restricted time range, now
showing a single molecular yo-yo cycle. The nucleosome unwinds at about 5.44~s and
 rewinds at about 6.02~s.
}
\label{Fig4}
\end{figure*}

\begin{figure}[h!]
\begin{center}
\includegraphics[trim=0cm 2.95cm 0cm 2.55cm, clip=true, width=3.1in]{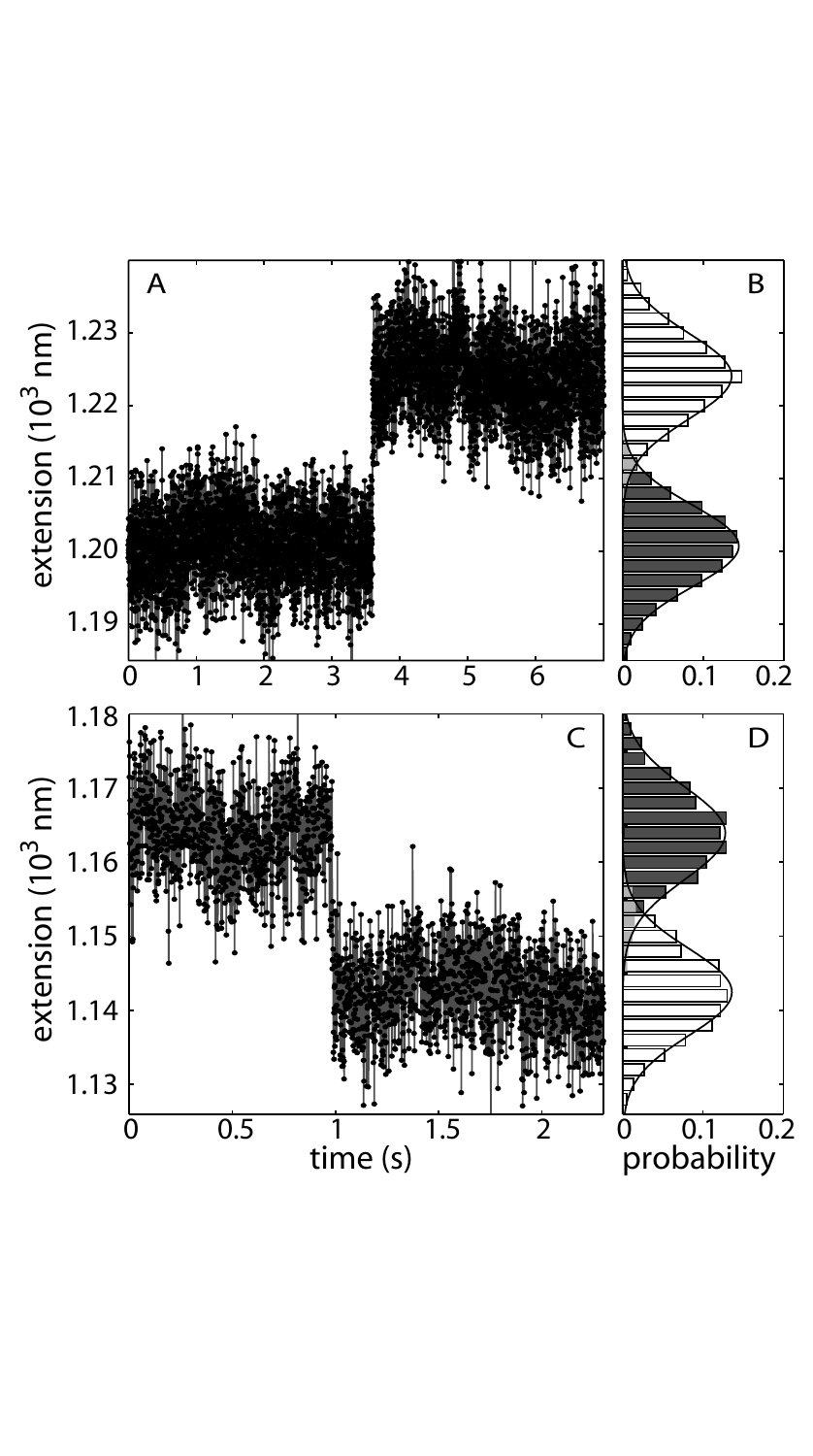}
\end{center}
\caption{Extension versus time of individual nucleosomes held at fixed force and their
corresponding extension distributions.
(A) Extension versus time of a single nucleosome, held at a fixed unwinding force of 10.3~pN,
measured every 1~ms and plotted as connected dots.
At 3.6~s after the unwinding force is applied, an unwinding event is evident, in which the extension jumps
by  25 nm.
(B) Experimental probability distributions for the extension before (gray bars)
and after (white bars) the unwinding event.
These histograms correspond to 3500 (gray) and 3300 (white) measurements of the extension with a bin size of 2~nm. 
The smooth curves are Gaussians, each plotted using the mean and standard deviation of the respective
experimental distributions:  $\sigma_1=5.9$~nm and $\sigma_0=5.6$~nm, for  state 1 and state 0, respectively, at this force.
(C) Extension versus time of a single nucleosome, held at a fixed rewinding force of 3.8~pN,
measured every 1~ms and plotted as connected dots. At 0.98~s after the force is applied,
a rewinding event
is observed as a stepwise 22~nm decrease in the extension.
(D) Probability distributions determined
before (gray bars) and after (white bars) the rewinding event with bins of size 2~nm.
These histograms correspond to 986 (gray) and 1314 (white) measurements of the extension. 
The smooth curves are Gaussians, each plotted using the mean and standard deviation of the respective
experimental distributions:  $\sigma_0=5.9$~nm and $\sigma_0=6.2$~nm, for  state 1 and state 0, respectively, at this force.}
\label{upJump}
\end{figure}

\section{RESULTS AND DISCUSSION}
\subsection{Molecular yo-yo method}
The molecular yo-yo method operates as follows:
At the start time, a high, unwinding force
is suddenly applied to the nucleosome in the wound state (state 1).
After a period of time at this force, the wound nucleosome transitions to the unwound state (state 0),
leading to an increase in extension.
By using the preprogrammed force-versus-extension curve of the states involved,
the yo-yo algorithm recognizes the change in state,
and, after a short delay,
reduces the force to the rewinding force.
After a period of time at the rewinding force, the nucleosome transitions back to the wound state (state 1),
with a concomitant decrease in extension.
The algorithm then recognizes this state change and increases the force to the unwinding force once again.
This cycle is then repeated multiple times, and at multiple unwinding and rewinding forces.
Importantly in the nucleosome context, 
the molecular yo-yo algorithm reduces as far as possible histone-DNA dissociation from the unwound state by minimizing the time the nucleosome spends in the unwound state.
Each measured dwell time at a given force contributes to the lifetime distribution at that force.

\begin{figure}[h]
\begin{center}
\includegraphics[trim= 0cm 0.75cm 0cm 0cm, clip=true, width=3.2in]{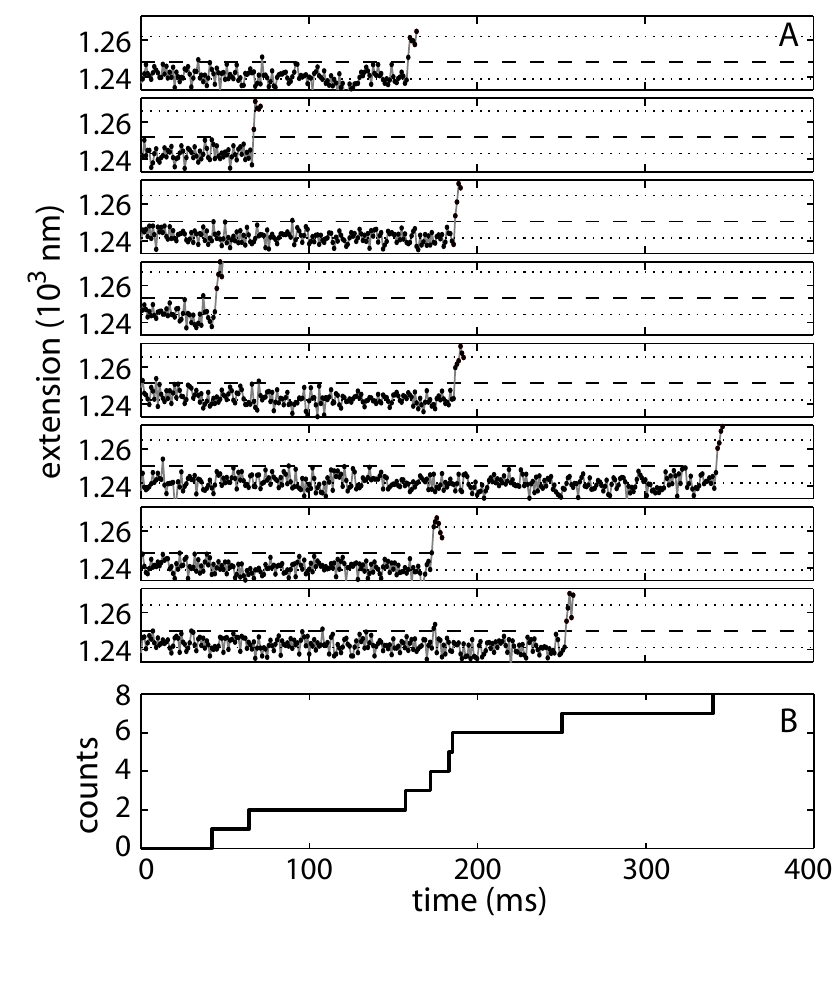}
\end{center}
\caption{Examples of individual nucleosome unwinding event detection. 
(A) Eight extension versus time traces at an unwinding force of 10.3~pN, each showing an example unwinding event recognized by our jump detection method. 
In each panel, the black dots connected by gray lines represent the extension measurement. The lower dotted line corresponds to the mean of the $n_1$ (40) points measured before the jump, and represents the extension of state 1, the wound state. 
The upper dotted line represents the extension of state 0, the unwound state, taken as the wound state’s extension plus 25~nm. 
The dashed line is 10~nm from the lower dotted line (state 1), and represents the threshold extension for jump detection in each trace. 
(B) Count of the number of unwinding events versus time, determined from the eight example traces shown in (A). 
}
\label{8upsFig}
\end{figure}

Fig.~\ref{Fig4}A shows a representative extension versus time trace
that covers twenty-eight repeats of the molecular yo-yo live jump detection algorithm that detects and triggers
on both unwinding and rewinding events.
Fig.~\ref{Fig4}B highlights nine of these cycles, and Fig.~\ref{Fig4}C a single cycle.
In Fig.~\ref{Fig4}C, for times prior to 5.25~s, the nucleosome inner turn is wound and the force is 3.8~pN.
At 5.25~s, the current yo-yo cycle is initiated, when the force-clamp steps up to 10.3~pN, as indicated by the green background. This force jump is
signaled by a corresponding stepwise increase in the extension trace of Fig.~\ref{Fig4}C, as the
DNA tether stretches further in response to the increased force.
The nucleosome remains wound at 10.3~pN for approximately 0.2~s until it unwinds at 5.44~s, signaled by a
jump in the extension of about 25~nm.
Thus, during this particular  cycle, the lifetime of the wound state at 10.3~pN was measured to be 0.2~s.
The nucleosome remains unwound at 10.3~pN for approximately 5~ms,
which is the period of time required for the yo-yo algorithm to recognize the unwinding event.
Once the transition to the
unwound state has been recognized, the force-clamp steps down to 3.8~pN, as indicated by the orange background.
The extension decreases concomitantly, because the DNA tether stretches less at the reduced force.
The nucleosome remains unwound at 3.8~pN until 6.01~s before rewinding.  This gives us a measurement
of the lifetime of the unwound state at 3.8~pN to be 0.55~s.
At 6.02~s, rewinding is signaled by a decrease in the extension of about 22~nm.
Once the rewinding transition is recognized, the
next yo-yo cycle is then initiated by
returning the force to 10.3~pN.  Triggering on both rewinding and unwinding events, as in Fig.~\ref{Fig4},
minimizes unnecessary measurements after the transition has occurred,
and leads to a significantly faster data acquisition rate than otherwise would be possible.

\subsection{Live jump detection algorithm}
We have implemented two jump detection algorithms, one for unwinding and one for rewinding.
Both jump detection algorithms rely on comparing live extension to preceding extension measurements. 
Fig.~\ref{upJump}A shows an example extension of a tether during a nucleosome unwinding event with no detection.
At time zero, a force-clamp of  10.3~pN is applied to a single
nucleosome.
For a period of approximately 3.6~s, the nucleosome in state 1 remains stable at this force, while the extension fluctuates about a
mean of about 1200~nm.
At 3.6~s,  the nucleosome unwinds into state 0, signaled by an increase in extension caused by about 25~nm of DNA being released.
Subsequently, state 0 is stable, with its extension fluctuating about  a mean of 1225~nm.
In Fig.~\ref{upJump}B, histograms of the measured extension, represented as probabilities, before (gray) and
after (white) the unwinding transition agree well with overlaid Gaussian distributions, plotted
using the mean and standard deviations of  the respective measured extension distributions.

Similarly, Fig.~\ref{upJump}C shows an example nucleosome rewinding event with no detection.
At time zero, a force-clamp of  3.8~pN is applied to a single
nucleosome, initially in state 0 (unwound).
For a period of approximately 0.98~s, the nucleosome in state 0 remains stable at this force, while the extension fluctuates about a mean of about 1163~nm.
At approximately 0.98~s,  the nucleosome rewinds into state 1, signaled by a decrease in extension of
about 22~nm.
Subsequently, state 1 is stable, with its extension fluctuating about  a mean of 1141~nm.
In Fig.~\ref{upJump}D, histograms of the measured extension, represented as probabilities, before (gray) and
after (white) the unwinding transition agree well with overlaid Gaussian distributions, plotted
using the mean and standard deviations of  the respective measured extension distributions.
These data were collected at a rate of 1~kHz, where extension fluctuations
effectively correspond to  uncorrelated, white noise.

To determine in real-time whether an unwinding transition has occurred,
our yo-yo algorithm looks back at the  previous $n_1+n_2$ extension measurements.
If all of the immediately preceding $n_1$ measurements exceed the mean of  the $n_2$
measurements previous to those $n_1$ measurements
by more than a threshold value ($\Delta_1$), then the yo-yo algorithm
recognizes that a jump in extension has occurred.
To assess  this simple scheme for false positives, we inquire:
What is the probability that the algorithm recognizes a transition from state 1 to state 0, when in fact none occurred?
 The probability that a single point exceeds the threshold, when the nucleosome remains in state 1,
 may readily be seen to be equal to
  \begin{equation}
P = \frac{1}{2} {\rm erfc}\left(\frac{\Delta_1}{ \sqrt{2 (1+\frac{1}{n_2})}\sigma_1}\right)
\simeq \frac{1}{2} {\rm erfc}\left(\frac{\Delta_1}{ \sqrt{2}\sigma_1}\right),
\label{erfc}
\end{equation}
 where $\rm erfc$ is the complementary error function,
 $\sigma_1$ is the standard deviation of the extension fluctuations in state 1,
 and
$\Delta_1$ is the difference in extension between the threshold extension
and the mean extension of state 1.
The factor $(1+1/n_2)$ accounts for the expected variance in the mean extension determined from $n_2$ measurements.
All of the unwinding yo-yo measurements presented in this paper employed $n_1=5$ and $n_2=40$.
With  $\sigma_1 = 5.9$~nm, as found experimentally at 10.3~pN,
it follows that $P=0.031$.
 Given the specified conditions for recognizing the transition,
 the probability of a false positive is $P^{n_1}$,
leading to a false negative probability of
$(0.031)^5=2.9\times10^{-8}$, which we consider entirely acceptable.
Similar considerations apply to unwinding at other forces.

\begin{figure}[h]
\begin{center}
\includegraphics[trim=0cm 0.5cm 0cm 0cm, clip=true, width=3.2in]{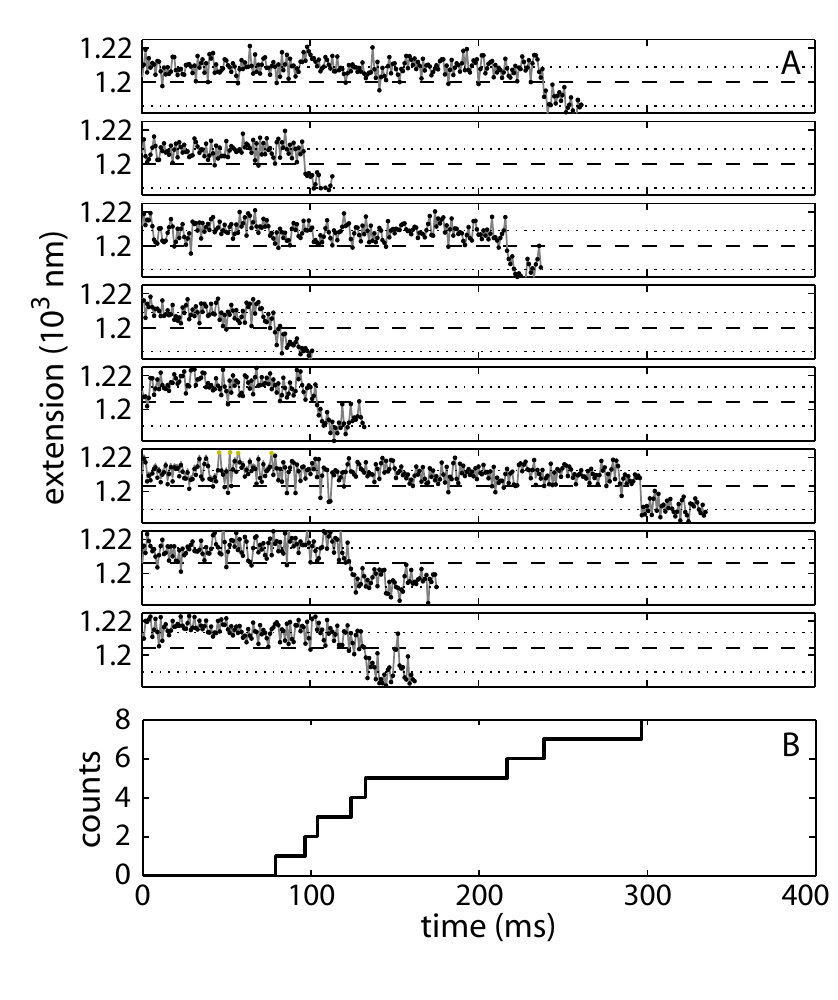}
\end{center}
\caption{
Examples of individual nucleosome rewinding event detection. 
(A)  Eight extension versus time traces at a rewinding force of 3.8~pN, each showing an example rewinding event recognized by our rewinding jump detection method. 
In each panel, the black dots connected by gray lines represent the extension measurement.
The lower dotted line
corresponds to the
previously-determined
extension of state 1
 at 3.8~pN.
The upper dotted line represents the extension of state 0,
taken as the extension of state 1 plus 23~nm. 
The dashed line,
10~nm from the lower dotted line (state 1),
represents the threshold extension for jump detection in these data. 
(B) Count of the number of rewinding events versus time, determined from the eight example traces shown in (A).
}
\label{8downsFig}
\end{figure}

Fig.~\ref{8upsFig}A
shows eight example nucleosome unwinding events, as detected by our jump detection algorithm.
In each case, the measurement begins with the nucleosome in the  state 1 (wound) and
with the application of the unwinding force of 10.3~pN at time zero.  After a variable time interval,
the nucleosome unwinds into state 0. In each case, after about 5~ms, corresponding to $n_1=5$, the jump detection algorithm
registers the unwinding transition and reduces the
force, having measured the lifetime of state 1 at the unwinding force.
In all cases, the transition is abrupt and is promptly detected by the algorithm.
In each panel, the lower dotted line corresponds to the
mean of the previous $n_2=40$ points before the jump,
and represents the extension corresponding to state 1.
The upper dotted line corresponds to the mean
extension of the unwound state at an extension 25~nm higher than the lower dotted line.
The dashed line represents the threshold, described previously,
and is located $\Delta_1=11$~nm above the mean extension of state 1.
Fig.~\ref{8upsFig}B summarizes the cumulative number of unwinding
events versus time, determined from the measurements of Fig.~\ref{8upsFig}A.

We use a modified jump detection algorithm for nucleosome rewinding. 
At the beginning of this protocol, the extension of a nucleosome in state~1 is determined by a 1~s extension measurement
at the rewinding force.
The extension of state 0 at the rewinding force is then taken to be the extension of state 1 plus 23~nm.
Once the force is jumped to the rewinding force,
our algorithm registers
a rewinding event when the mean of the previous $n_1$ extension measurements lies within a threshold value, $\Delta_1$,
of the extension of state~1. 
In this case, the probability of falsely identifying a transition, when none has occurred, is
\begin{equation}
P 
= \frac{1}{2} {\rm erfc}\left(\frac{\Delta_1\sqrt{n_1}}{ \sqrt{2}\sigma_1}\right).
\label{erfc1}
\end{equation}
Fig.~\ref{8downsFig}A shows eight example nucleosome rewinding events, as detected by this algorithm.
In each case, the measurement begins with the nucleosome in the
state 0 and
with the application of the rewinding force of 3.8~pN at time zero.  After a variable time interval,
the nucleosome rewinds into state 1. Then, after a time 
that can be seen to vary from 10 to 40 ms, the jump detection algorithm
registers the rewinding transition and increases the
force, having measured the lifetime of state 0 at the rewinding force.
In all cases, the transition is unambiguous and is readily detected by the algorithm,
although not as rapidly as in the case of unwinding.
For these example rewinding events, for which  $\sigma = 6~$nm,
 $n_1= 40$,  and $\Delta_1=10$~nm, a false-positive jump detection occurs with a
 probability of $3\times10^{-29}$, effectively impossible on experimental time-scales.
Fig.~\ref{8downsFig}B summarizes the cumulative number of rewinding versus time,
determined from the measurements of Fig.~\ref{8downsFig}A.

\begin{figure}[t]
\begin{center}
\includegraphics[trim=0.0cm 2.6cm 0.0cm 1.95cm, clip=true]{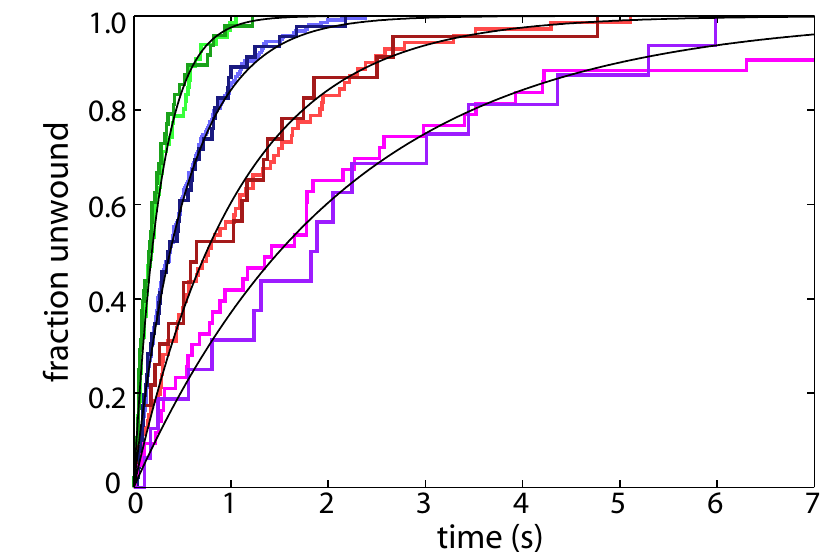}
\end{center}
\caption{Cumulative fraction of nucleosomes unwound versus time at 10.3~pN determined on tethers with 8~(green), 4~(blue), 2~(red), or 1~(purple) nucleosomes.
Each curve corresponds to measurements performed on a single tether.
Each step in the staircases corresponds to an independent unwinding event.
The smooth curves are exponentials with
rates $8k$, $4k$, $2k$, and $k$, where $k$ is determined by the maximum likelihood method from the single nucleosome data \cite{Mack2012a}. The apparent rate scales with the number of nucleosomes.
Measurements on different tethers give the same rate for single nucleosome unwinding.
}
\label{linearFig}
\end{figure}

\subsection{Distribution of lifetimes at fixed force}

To determine the lifetime distribution of state 1 or state 0 at each force
studied, we count the cumulative number of unwinding or rewinding events as
a function of time, as indicated in Fig.~\ref{8upsFig}B and Fig.~\ref{8downsFig}B.
Normalizing by the total number of counts,
these collections of unwinding and rewinding
times determine the cumulative fraction of nucleosome inner turns unwound and wound, respectively,
as a function of time, at a given force.
In Fig.~\ref{linearFig},
we show unwinding data at 10.3~pN obtained on tethers with one, two, four and eight nucleosomes in purple, red, blue and green respectively.  
We plot two curves each for 1, 2, 4, and 8 nucleosomes. Each was obtained in separate experiments with different tethers and nucleosomes, demonstrating individual tether-to-tether and nucleosome-to-nucleosome repeatability.

If we assume first-order unwinding (rewinding) kinetics,
 the probability that a nucleosome initially in state~1 (or 0)
will have undergone unwinding to state~0 (or 1)
within a time $t$ after initiation of the force clamp is given by
an exponential function:
\begin{equation}
p = 1 - e^{-kt},
\label{FRACTIONUNWOUND}
\end{equation}
where $k$ is the unwinding (rewinding) rate.
For independent nucleosomes, we expect the apparent unwinding rate for 2, 4, and 8 nucleosomes to be 2 times, 4 times  and 8 times, respectively, the rate for a single nucleosome.
The smooth curves in Fig.~\ref{linearFig} correspond to exponentials with rate set according to this rule,
using the maximum likelihood value of the unwinding rate determined from data obtained on a single nucleosome.
Evidently,  the model curves provide an excellent description of  the behavior with 1,  2, 4, and 8 nucleosomes.
This observation directly demonstrates that at the forces studied nucleosomes on the same DNA tether unwind independently,
as has previously been assumed but not proven.

\begin{figure}[t]
\begin{center}
\includegraphics[trim=0cm .15cm 0cm 0.0cm, clip=true, width=3.2in]{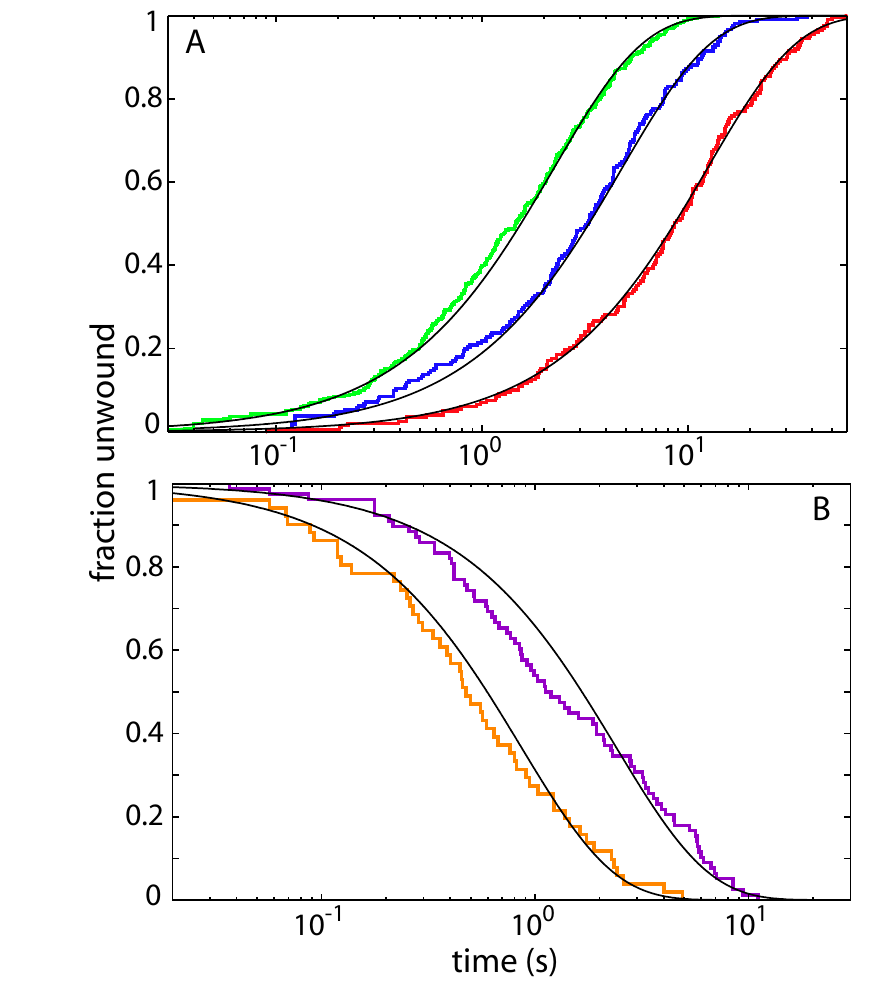}
\end{center}
\caption{Cumulative fraction of nucleosomes unwound versus time for unwinding and rewinding determined using the molecular yo-yo method. (A) Distribution of unwinding times at 8.5~pN (red), 9.4~pN (green), and 10.3~pN (blue) shown using a logarithmic time axis, with disitrubutions determined using 201, 211,  and 354 unwinding events, respectively. Each step is an independent unwinding event.
(B) Distributions of rewinding times at 3.3~pN (orange) and 3.8~pN (purple), after unwinding at 14.1~pN,
determined using  66 and 78 rewinding events, respectively. 
In both (A) and (B), the corresponding, overlayed smooth black curves are exponentials with rates determined by the maximum likelihood method.}
\label{semilogFig}
\end{figure}

Fig.~\ref{semilogFig}A displays the distributions of unwinding times,
using a logarithmic time axis,
 at 8.5, 9.4, and 10.3~pN,
each determined
from multiple nucleosomes, collected together to yield a single distribution at each force.
 Similarly, Fig.~\ref{semilogFig}B displays the distribution of rewinding times at 3.3 and 3.8~pN
for nucleosomes, that were unwound at 14.1~pN.
For both Fig.~\ref{semilogFig}A and B,
each distribution is represented as the fraction of nucleosomes unwound.
The collected events at each force
constitute a sufficiently large data set (201, 211,  354, 66 and 78 transitions at 8.5, 9.4, 10.3, 3.3 and 3.8~pN, respectively)
to enable us to not
only determine the unwinding rate at the force in question but also to
test whether an exponential distribution of lifetimes is a correct description.
The solid black lines in Fig.~\ref{semilogFig}A and B
correspond to EQ. \ref{FRACTIONUNWOUND} calculated using the
maximum likelihood values of the unwinding and rewinding rates respectively.
Evidently,  this model provides an excellent description of our measured
lifetime distributions with zero adjustable parameters,
indicating that a single exponential lifetime distribution is the correct
description, and that the transition rates are
$0.062 \pm 0.004$, $0.16  \pm 0.01$, $0.40\pm 0.02$,  $1.2 \pm 0.1$, and $0.38 \pm 0.04~s^{-1}$ at 8.5, 9.4, 10.3, 3.3, and 3.8~pN, respectively.
To objectively assess how well
EQ. \ref{FRACTIONUNWOUND} accounts for the measured lifetime distributions,
we have binned the unwinding lifetime measurements shown in Fig.~\ref{semilogFig}
into  logarithmically-sized bins \cite{Sigworth1987}.   
Binning the data to obtain the distribution of lifetimes
ensures that the number of counts in the different
bins are statistically independent of one another,
which is not the case for the cumulative distributions of
Figs.~\ref{linearFig} and \ref{semilogFig}.
It also permits us to simply determine the standard
error for each bin as the square-root of the number of counts in each bin.
The corresponding histograms, including error bars, are compared with
the model distribution corresponding to EQ.~\ref{FRACTIONUNWOUND}
in Fig.~\ref{histogramFig}. 
To determine the goodness of fit, we calculated the reduced chi-squared:
\begin{equation}
\chi^2 =\frac{1}{n-1}{\Large  \Sigma}_{i=1}^{n} \frac{(O_i-E_i)^2}{E_i^2},
\end{equation}
where the sum runs from 1 to $n=13$  bins, $O_i$ is the observed number of counts in
 bin $i$, and  $E_i$ is the expected number of counts in bin $i$ \cite{taylor1997introduction}.
The reduced $\chi^2$-values are 0.58, 1.3, and 1.21,  for the
unwinding distributions at 8.5, 9.4, and 10.3~pN, 
respectively.
For the rewinding distribution at 3.3 and 3.8~pN,
the reduced $\chi^2$-values are  1.0 and 1.3, respectively.
Thus, in every case, $\chi^2$ is close to unity,
indicating that there
is not a statistically significant deviation between the data and the model
of EQ.~\ref{FRACTIONUNWOUND} at any of the forces studied.

\begin{figure}[t]
\begin{center}
\includegraphics[trim=0cm .2cm 0cm 0.2cm, clip=true, width=3.2in]{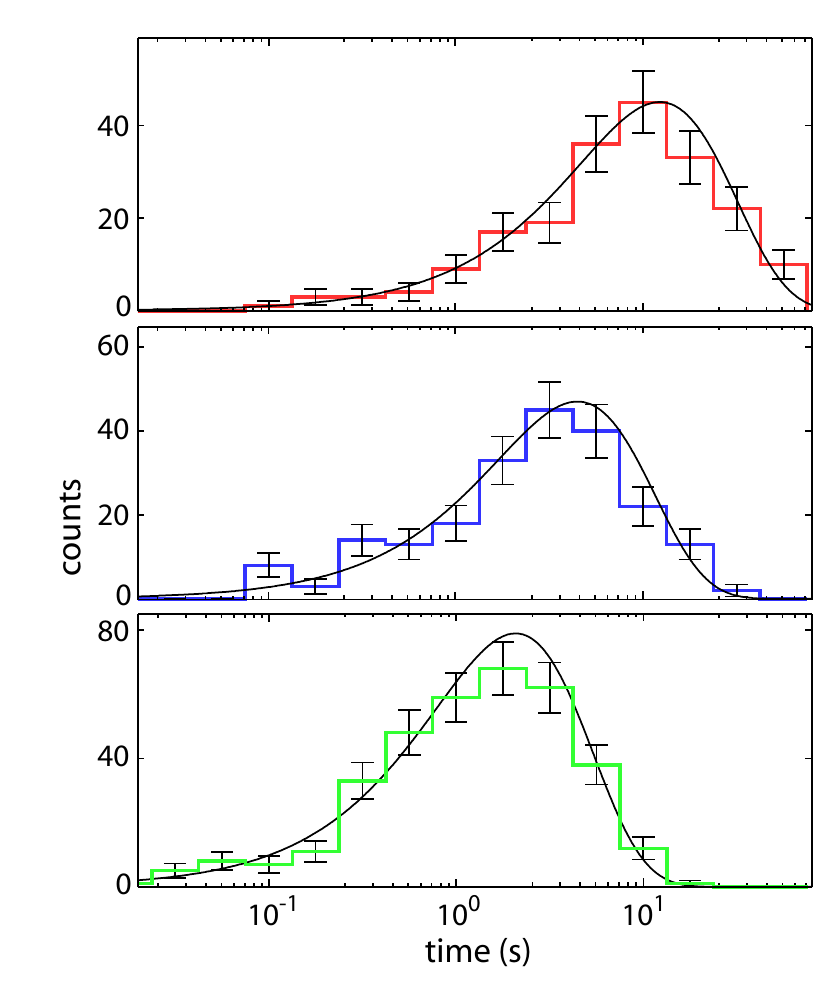}
\end{center}
\caption{Histogram of nucleosome unwinding times shown using a logarithmic time axis. 
Unwinding times at 8.5~pN (red), 9.4~pN (blue), and 10.3~pN (green) were logarithmically-binned 
with distributions determined using 201, 211, and 354 unwinding events, respectively.
Each bin is a factor of 3 longer than the preceding bin.
The histogram of the number of events with lifetimes in each time
bin is displayed as a staircase.
The standard error for the number of counts in each bin, shown as the error bar,
is determined by counting (Poisson) statistics.
An exponential distribution, corresponding to the characteristic
unwinding rate,  determined by the maximum likelihood method,
is overlaid as the thin continuous line.
}
\label{histogramFig}
\end{figure}

\subsection{Force dependent unwinding and rewinding rates of the
nucleosome inner turn}
Fig.~\ref{rateFig} summarizes our measurements of
the force-dependent rates of unwinding
and  rewinding the nucleosome inner turn,
obtained with the molecular yo-yo method.
For  comparison, Fig.~\ref{rateFig}
also includes our previously published results
for these rates
\cite{Mack2012a}.
Evidently, measurements of the nucleosome inner turn unwinding and rewinding rates, obtained using the molecular yo-yo method,
show good agreement with those obtained previously.
At a given force, however, the number of transitions available via the yo-yo method is
several-fold larger than the number previously available,
yielding  values for the transition rates that are accurate to within about 10\%.
Thus, in future nucleosome studies the molecular yo-yo method will enable us to resolve subtle
differences in the kinetics of different nucleosomes,  with different histone variants and modifications, for example.

\begin{figure}[h]
\begin{center}
\includegraphics[trim=0cm 2cm 0cm 2.0cm, clip=true, width=3.2in]{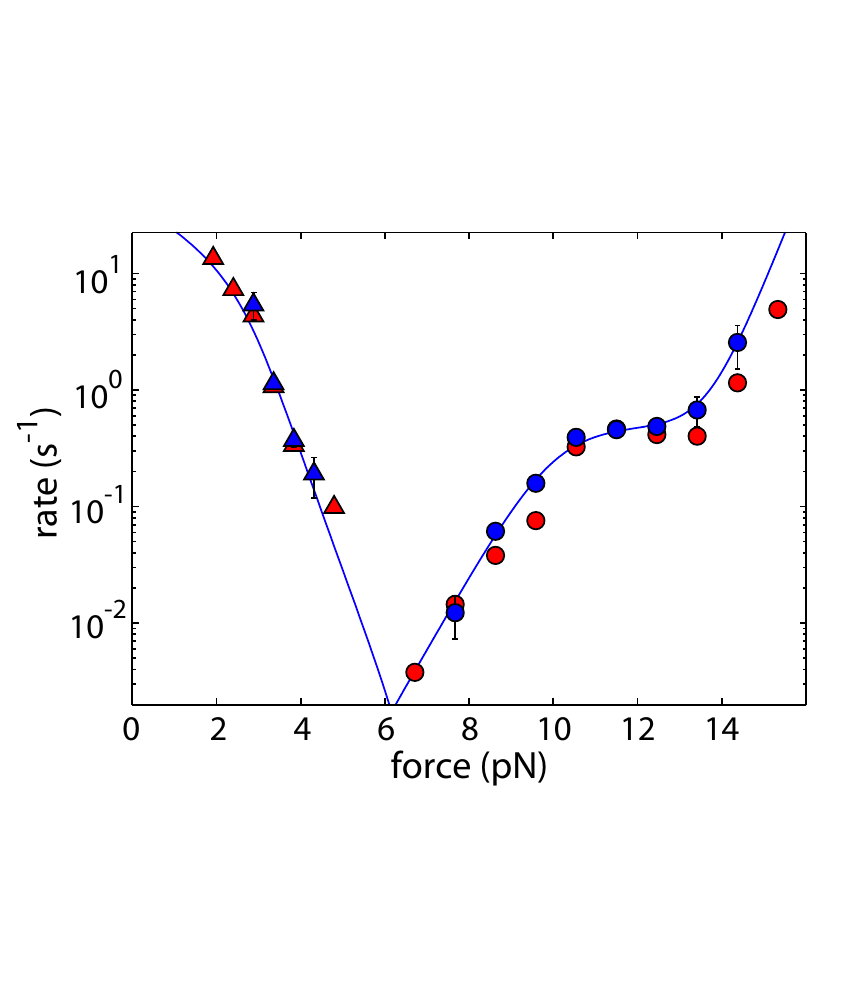}
\end{center}
\caption{Unwinding and rewinding rates of the nucleosome inner turn plotted
as a function of force.
Shown as blue circles are the unwinding rates obtained from 12 nucleosomes
undergoing a total of 1086 unwinding events
with the molecular yo-yo method.
Shown as blue triangles are the rewinding rates, obtained using the molecular yo-yo method, from 6 nucleosomes undergoing a total of 167 rewinding events.
The previously published
unwinding and rewinding rates are shown as red  circles and triangles, respectively.
These data were acquired using a total of 334 nucleosomes with 265 unwinding events and 136 rewinding events.
}
\label{rateFig}
\end{figure}

\subsection{Conclusion}
In this paper, we have introduced and demonstrated a powerful elaboration of
the force clamp method, called the molecular yo-yo, that is broadly applicable to molecular transitions that are far from equilibrium.
The molecular yo-yo method implements a live jump-detection and force-clamp algorithm, that intelligently adjusts and maintains the force on a single molecule, in response to the measured state of that molecule.
Thus, we are able to 
 realize hundreds of individual molecular 
transitions between molecular states at 
 different forces,
 permitting us to accurately determine force-dependent lifetime distributions and reaction rates.
 Compared to force-versus-extension measurements, 
the  molecular yo-yo method directly measures these key quantities,
while maximizing data acquisition rate and efficiency.
Compared to force-jump measurements, the molecular yo-yo method
minimizes the time spent at high force, where molecular complexes can dissociate.
We presented measurements detailing the specific application of the molecular yo-yo to
unwinding and rewinding the nucleosome inner turn, using optical tweezers.
Because the molecular yo-yo minimizes the time spent in the unwound state (state 0),
nucleosome dissociation is also minimized, permitting hundreds of transitions to be
obtained from a single construct before nucleosome dissociation ends the measurement.
Our molecular yo-yo measurements of unwinding and rewinding the nucleosome
inner turn reveal experimental lifetime distributions  that are accurately single exponential,
indicating the existence of a single dominant free energy barrier between states 1 and 0.
We also demonstrate that the unwinding rates for tethers containing 2, 4, and 8 nucleosomes are accurately 2-, 4-, and 8-fold faster, respectively, than for tethers containing a single nucleosome.
This observation implies that nucleosomes on the same tether unwind independently,
as has been previously assumed but not proven.
Finally, we note that improved  throughput is not the only benefit allotted by live jump detection,
 only the most obvious.
 A variety of more elaborate experimental force protocols
 enabling new measurements are clearly made possible
 with variations of the technique presented here.
 Additionally, live jump detection methods are transferable to other SMFS methods, such as those that employ an atomic force microscope.

\begin{acknowledgments}
We thank J. Antonypillai, E. Dufresne, R. Illagan, P. Koo, L.-A. Metskas,
 F. Sigworth, Y. Zhao, N. Sawyer, E, Speltz, A. Schloss, J. Chen, and A. Zhou  for valuable discussions.
This work was supported by the Raymond and Beverly Sackler Institute for Biological, Physical and Engineering Sciences
and NSF PoLS 1019147.
D.J.S. acknowledges the support of a NSF Graduate Research Fellowship.
M. K. acknowledges the support of NSF Postdoctoral Research Fellowship in Biology Award  DBI 1103715.
\end{acknowledgments}

\bibliographystyle{apsrev}

\end{document}